# Two distinct kinetic regimes for the relaxation of light-induced superconductivity in $La_{1.675}Eu_{0.2}Sr_{0.125}CuO_4$

C. R. Hunt[1,2], D. Nicoletti[1], S. Kaiser[1], T. Takayama[3,4], H. Takagi[3,4,5], and A. Cavalleri[1,6]

[1] *Max Planck Institute for the Structure and Dynamics of Matter, Hamburg, Germany*

[2] *Department of Physics, University of Illinois at Urbana-Champaign, Urbana, Illinois, USA*

[3] *Department of Advanced Materials Science, University of Tokyo, Tokyo, Japan.*

[4] *Max-Planck-Institut für Festkörperforschung, Heisenbergstrasse 1, D-70569 Stuttgart, Germany*

[5] *RIKEN Advanced Science Institute, Hirosawa 2-1, Wako 351-0198, Japan*

[6] *Department of Physics, Oxford University, Clarendon Laboratory, Oxford, United Kingdom*

**We address the kinetic competition between charge striped order and superconductivity in $La_{1.675}Eu_{0.2}Sr_{0.125}CuO_4$. Ultrafast optical excitation is tuned to a mid-infrared vibrational resonance that destroys charge order and promptly establishes transient coherent interlayer coupling in this material. This effect is evidenced by the appearance of a longitudinal plasma mode reminiscent of a Josephson plasma resonance. We find that coherent interlayer coupling can be generated up to the charge order transition $T_{CO} \approx 80$ K, far above the equilibrium superconducting transition temperature of any lanthanide cuprate. Two key observations are extracted from the relaxation kinetics of the interlayer coupling. Firstly, the plasma mode relaxes through a collapse of its coherence length and not its density. Secondly, two distinct kinetic regimes are observed for this relaxation, above and below spin order transition $T_{SO} \approx 25$ K. Especially, the temperature independent relaxation rate observed below $T_{SO}$ is anomalous and suggests coexistence of superconductivity and stripes rather than competition. Both observations support arguments that a low temperature coherent stripe (or pair density wave) phase suppresses c-axis tunneling by disruptive interference rather than by depleting the condensate.**

Stripe order in cuprates is closely associated with the suppression of superconductivity, although the interplay between these orders remains a subject of much debate. In these materials, superconductivity is thought to be supported in two-dimensional $CuO_2$ planes and made three-dimensional by interlayer Josephson tunneling. However both in-plane and out-of-plane coherence can be strongly affected by ordering of charges and spins, or by lattice deformations. Small perturbations in doping, applied field, or pressure can tune the energy landscape between orders, suppressing or supporting the superconducting state.

This phase competition is especially dramatic in the lanthanum copper oxides, which exhibit "striped" spin and charge ordered states, typically stabilized by an underlying lattice distortion[1,2]. The charge and spin orders also organize within the $CuO_2$ planes. Static stripe orders, first discovered in $La_{1.6-x}Nd_{0.4}Sr_xCuO_4$ (LNSCO $x$)[3], are also found in $La_{2-x}Ba_xCuO_4$ (LBCO $x$)[4,5] and $La_{1.8-x}Eu_{0.2}Sr_xCuO_4$ (LESCO $x$)[6,7] (see phase diagram in figure 1A). Charge stripes are associated with the suppression of superconductivity in this family of compounds, with bulk superconductivity completely destroyed at $x = 1/8$ doping[8,9] where the lattice spacing synchronizes with the stripe periodicity.

Recent theoretical[10,11,12,13] and experimental[9,14,15] work suggests that this suppression is not a simple competition between charge density wave and superconducting instabilities. Rather, superconductivity may coexist with charge ordering, which imposes a space dependence on the superconducting order parameter phase[11]. This spatial modulation—often referred to as a "pair density wave" state—suppresses the total Josephson tunneling by disruptive interference[12]. According to this view, 3D superconductivity would only be achieved below the temperature at which the superconducting $c$-axis coherence length exceeds the quadrupled unit cell spacing[16] (see inset in figure 1A). This assessment is supported by susceptibility[9] and resistivity and thermopower[15] measurements that suggest a fluctuating 2D superconductivity regime survives in the spin order state, up to $T_{SO} = 40$ K in LBCO 1/8, and 1D correlations persist up to the charge order transition, $T_{CO} = 54$ K.

In this work, we used femtosecond laser excitation to reintroduce c-axis coherent coupling in stripe ordered $La_{1.675}Eu_{0.2}Sr_{0.125}CuO_4$ (LESCO 1/8) and examined the relaxation kinetics. LESCO 1/8 single crystals were grown using the traveling solvent floating zone technique. They were characterized by resistivity and magnetization measurements and were found to be non-superconducting down to 5 K.[17] The charge- and spin-ordered regimes for this doping are indicated by the dashed vertical line on the phase diagram in Figure 1A.

Laser excitation has proven a powerful tool to control lattice[18,19,20,21] and electronic[22,23,24,25] properties and drive phase transitions. The 15 $\mu$m mid-infrared (MIR) pump used here, tuned to an in-plane Cu-O mode, has been shown to promptly reduce charge stripe order[26] and reintroduce c-axis coherent coupling[27]. The c-axis optical response below 2.6 THz was interrogated as a function of time delay after photo-excitation. Single-cycle THz pulses were generated via 800 nm excitation of a photoconductive antenna. The time resolution of the experiment was limited by the THz bandwidth to about 300 fs. The experiment was performed in reflection geometry, with the MIR pump at normal incidence and the THz *s*-polarized along the *c*-axis at a 30° angle of incidence. The reflected equilibrium THz field, $E(t)$, and the pump-induced changes to the field at time delay $\tau$, $\Delta E(t,\tau)$, were measured via electro-optic sampling to capture the full amplitude and phase response.

The raw change in reflectivity, given by the Fourier transformed quantity $|\Delta \widetilde{E}(\omega,\tau)/\widetilde{E}(\omega)| \equiv \Delta R/R$, is shown in figure 1B at $\tau = 1.8$ ps after excitation, which corresponds to the peak amplitude of the transient response. Below the charge order transition temperature, the transient spectra show the appearance of a reflectivity edge between 1.5 and 2 THz. Figure 1B.1 shows 5 K base temperature and 1B.2 shows 30 K, above $T_{SO}$. Note that the appearance of this effect has been reported in Ref. 27 up to 10-20 K. Improvements in the experimental apparatus and a fourfold increase in the excitation fluence (to ~4 mJ/cm²) make it possible here to detect a qualitatively similar but smaller edge far above $T_{SO}$, up to between 70 to 80 K (dark grey dots, figure 1C). Above $T_{CO}$, the THz response shows no reflectivity edge (figure 2B.3 at 100 K).

The complex optical properties of the photo-induced state were calculated from the THz response and equilibrium properties measured by Fourier Transform Infrared (FTIR) spectroscopy. The THz probe samples a crystal volume on order 100 times greater than the 15 $\mu$m pump. To isolate the optical response of the photo-excited volume alone we model the system as a single excited layer on a bulk volume that remains in equilibrium[28] (see supplementary).

Figure 2 shows the optical response of LESCO 1/8 at two temperatures, below and above the spin order transition $T_{SO} \approx 25$ K.[6,29] In both regimes, the transient response (dots) is characterized by the appearance of a longitudinal plasma mode at ~1 THz. This mode is most clearly discerned in the real dielectric response $\varepsilon_1(\omega)$, shown in figure 2A, which exhibits a zero crossing near the mode resonance. The Ohmic conductivity $\sigma_1(\omega)$ is only weakly affected by the pump, maintaining an insulator-like response (figure 2B), suggesting the pump is not producing significant quasiparticle excitations. The inductive conductivity, $\sigma_2(\omega)$, which is negative in equilibrium (grey line) and approaches zero as $\omega \to 0$, instead increases divergently towards low frequency in the photo-excited state (figure 2C). The response turns positive below 0.75 THz at 5 K and 0.5 THz at 30 K. The conductivity change, $\Delta\sigma_2(\omega) = \sigma_2(\omega) - \sigma_{2,eq}(\omega)$, scales as $1/\omega$, as shown by the dotted line in figure 2D.

In equilibrium, LESCO 1/8 has no plasma mode in this frequency range at any temperature. However, the transient mode coincides with the frequency of the Josephson plasma resonance (JPR) of related underdoped superconducting lanthanides[14,30]. A Josephson plasma mode is a generic feature of superconductivity in cuprates which arises due to the tunneling of pairs between CuO$_2$ planes. Its frequency $\omega_p$ is related to the condensate density, compressibility, and the geometric spacing of the planes.[31] Within one family of compounds, the JPR frequency scales with the superfluid density as $\omega_{JPR}^2 \propto n_{SF}$, implying that the transient state reported here has a condensate density of roughly half that of near-optimal doped La$_{1.85}$Sr$_{0.15}$CuO$_4$ in equilibrium.

Hence, although the appearance of a longitudinal plasma mode by itself does not uniquely prove superconductivity, the quantitative agreement with the optical response of related superconducting compounds makes the assignment of a superconducting state by far the most likely explanation.

The solid black lines in Figure 2 show a fit to a single longitudinal plasma mode utilizing only two free parameters[32] with the Drude form,

$$\tilde{\varepsilon} = \tilde{\varepsilon}_{eq} - \frac{\omega_p^2}{\omega^2 - i\omega\Gamma},$$

where $\tilde{\varepsilon}_{eq}$ is the equilibrium dielectric function. The transient response is best fit by $\omega_p = 2.45$ THz (1.65 THz) at 5 K (30 K) with a scattering rate of $\Gamma \approx 0.25$ THz.

Within this model, the scattering rate term $\Gamma$ encompasses all transient processes that impact the mobility along the c-axis. The scattering rate is extremely low, and cannot be reconciled with the properties of an incoherent plasma, but is consistent with a superconducting state with finite coherence length such as that seen in the equilibrium superconductor La$_{2-x}$Sr$_x$CuO$_4$ near $T_c$[33]. At resonance, transport occurs across the CuO$_2$ planes with a velocity $2\omega_P L$, where $L$ is the CuO$_2$ plane separation. The rate $\Gamma$ can be related to the coherence length $d$ of the c-axis plasma by $d = 2\omega_P L/\Gamma$.

Note that the reflectivity edge and the zero crossing of $\varepsilon_1(\omega)$ do not appear exactly at the plasma resonance, $\omega_P$, but are shifted due to decoherence as well as other intra-band contributions to $\varepsilon_{1,eq}$ which can be captured in the THz regime by a single parameter $\varepsilon_{\text{FIR}}$. For long coherence lengths, $d \to \infty$, the zero crossing occurs near the screened frequency $\tilde{\omega}_P = \omega_P/\sqrt{\varepsilon_{\text{FIR}}}$ and shifts to the red as $d$ decreases. If $\Gamma > \tilde{\omega}_P$, the zero crossing is entirely lifted. For the plasma mode $\tilde{\omega}_P$ reported here, we take $\varepsilon_{\text{FIR}} = 30$, a standard value for cuprates.[34,35]

Figure 3 shows the evolution of $\varepsilon_1(\omega)$ as a function of delay time after photo-excitation at 5 K (A.1) 35 K (A.2) and 65 K (A.3). From fits to the optical response, we extract the time evolution of the screened plasma mode $\tilde{\omega}_P$ (figure 3B, bottom) and the coherence length $d$ (figure 3B, top). After the transient state is formed, both quantities initially decay following a 2 ps (1 ps) timescale at 5 K (35 K). At longer time delays, the plasma frequency stabilizes to a finite value, indicating that the carrier density does not reduce significantly throughout the relaxation. This observation offers further support that the transient state is not consistent with an anomalously high-mobility quasiparticle excitation, which would relax via a depletion of the carrier density. Rather, the decay of the plasma mode is characterized by a dramatic decrease in the correlation length from ~15 unit cells (10 nm) to zero.

The highest temperature at which the longitudinal mode could be seen in $\varepsilon_1(\omega)$ in our frequency window was 65 K. The low-frequency cut-off is limited by the sample size and by day-to-day alignment to between 0.2-0.4 THz. From the optical response, and the amplitude of the reflectivity changes $\Delta E(\tau)/E^{max}$, we extrapolate that the maximum base temperature at which the transient state can be induced is between 70 K and 80 K, which is approximately the charge ordering temperature $T_{CO}$.

The relaxation timescales are also captured by the changes in reflectivity measured at the peak of the THz field, $\Delta E(\tau)/E^{max}$, which could be measured with finer delay steps. Figure 4A shows $\Delta E(\tau)/E^{max}$ as a function of pump-probe delay $\tau$. Two lifetimes could be extracted, shown in blue in figure 4B. The lifetimes measured from the coherence length decay are shown in red for comparison. In both the spin-ordered and charge-ordered regimes, the time scales of the double exponential decay are commensurate with the decay of the coherence length. Both lifetimes exhibit two distinct regimes. Below $T_{SO}$, the lifetimes remain temperature independent. Above $T_{SO}$, where only static charge order remains, the lifetime drops exponentially with base temperature.

The exponential dependence of the relaxation between $T_{SO} < T < T_{CO}$ can be reconciled with the expected kinetic behavior for a transition between two distinct thermodynamic phases separated by a free energy barrier. This is quantitatively captured by the slope of the logarithmic plot in figure 4B, which reflects an activated relaxation of the type $\exp(-E_{barrier}/k_B T)$. From a double exponential relaxation with lifetimes $\tau_1$ and $\tau_2$, we extract an energy scale $E_{barrier} \sim 40$ K (4 meV) and 9 K (0.8 meV) respectively. Interestingly, this energetic regime corresponds to the energy scale of spin fluctuations measured in LSCO 1/8,[36,37,38] suggesting that the transition between the two phases may be regulated by spin rearrangements. This is consistent with the observation that strong spin fluctuations develop in equilibrium above $T_{SO}$ and survive up to $T_{CO}$.[39,40]

The departure from activated behavior for $T < T_{SO}$ may therefore be related to the freezing out of spin fluctuations. A temperature-independent relaxation rate would be compatible with quantum coherent tunneling between two states, for instance between superconducting and pair density wave phases at constant carrier density. In this picture, Cooper pairing would be superimposed or intertwined[41,42,43] with the stripe phase, where the dynamical destruction of stripes allows a finite Josephson current rather than driving pairing directly. Indeed, as discussed in Ref. 27, the prompt timescale of the appearance of the longitudinal mode renders it unlikely that the optical excitation is causing pair formation[44] but rather suggests that pairing in the planes persists in equilibrium.

We have reported the generation of a transient high mobility state in LESCO 1/8 that survives up to the charge order transition. This transient state is characterized by the appearance of a longitudinal plasma mode in the low frequency THz response, which we attribute to *c*-axis superconducting coupling. A first striking observation is that the transient plasma mode can be induced all the way up to the charge order transition temperature $T_{CO}$ ~ 80 K, a large temperature for this family of cuprates. This observation further substantiates theoretical ideas that charge ordered cuprates remain frustrated superconductors above their equilibrium transition temperature, and that pairing persists even above the spin ordering temperature $T_{SO}$. We find that two distinct regimes of phase competition exist between stripe order and transient superconductivity. The lifetime of the photo-induced state remains essentially independent of base temperature below $T_{SO}$ and follows an Arrhenius-like behavior in the charge ordered regime, up to $T_{CO}$. Our measurements support the growing body of evidence for 2D superconducting pairing in the charge ordered state of cuprates.

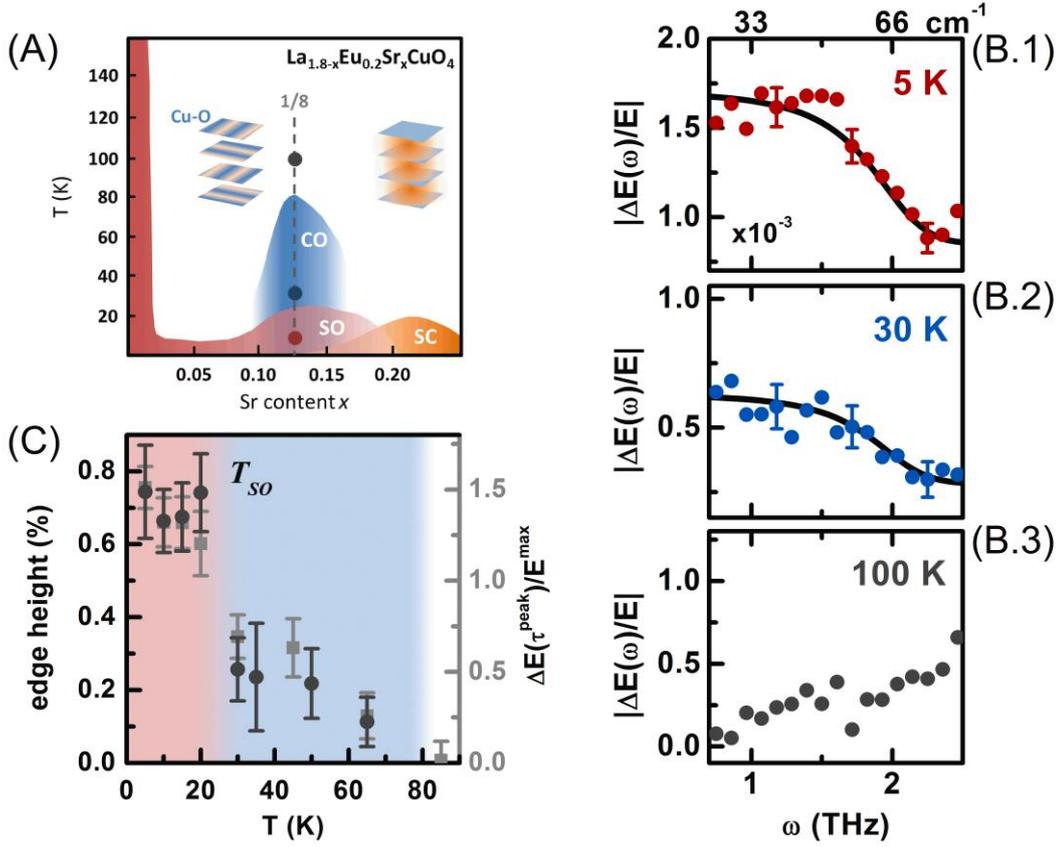

**Figure 1. Phase diagram of LESCO and transient response at 5 K, 30 K, and 100 K. (A)** Phase diagram of LESCO, based on Ref. 29 indicating regions of bulk superconductivity (SC), static spin (SO) and charge (CO) order. The static stripes suppress $c$-axis coupling of the $CuO_2$ planes (inset cartoon, left), with bulk superconductivity restored at dopings in which the stripe order is reduced (inset cartoon, right). **(B)** The raw transient reflectivity changes measured 1.8 ps after MIR excitation. At 5 K **(B.1)** and 30 K **(B.2)**, an edge is apparent at 1.5-2 THz. The black lines indicate the reflectivity spectrum (rescaled) due to a longitudinal plasma mode, shown as a guide to the eye. Above the charge ordered transition temperature, $T_{CO}$, no edge is observed (shown at 100 K, **B.3**). **(C)** The size of the reflectivity edge remains approximately constant below the spin order transition temperature $T_{SO}$, but drops rapidly between $T_{SO}$ and $T_{CO}$ (dark grey circles). This trend is also reflected in the time domain, by the change in the THz amplitude $\Delta E(\tau)/E^{max}$ measured at the peak of the response (light grey squares). The full delay traces $\Delta E(\tau)/E^{max}$ are shown in figure 4.

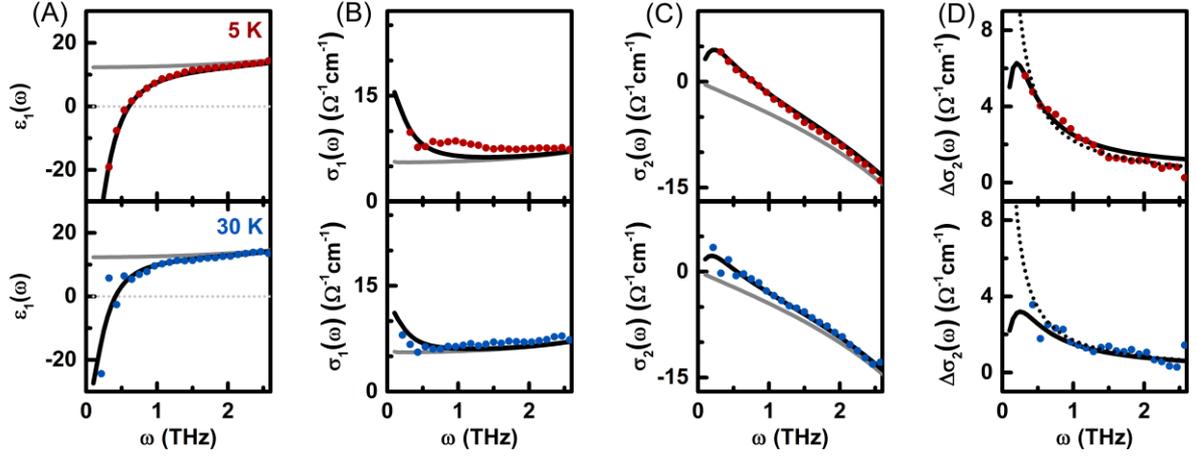

**Figure 2. Transient optical properties at 5 K and 30 K.** Upon MIR excitation, LESCO 1/8 develops a high mobility state, shown 2.8 ps after excitation at 5 K (red, first row) and at the peak of the response at 30 K (blue, second row). The equilibrium response is shown in grey. The transient state can be fit (black lines) by a longitudinal plasma mode, as described in the main text. **(A)** The mode is characterized by a zero crossing in the real dielectric function, $\varepsilon_1(\omega)$. **(B)** The Ohmic conductivity, $\sigma_1(\omega)$, remains insulating, showing only a small enhancement at lowest frequencies. **(C)** The inductive conductivity, $\sigma_2(\omega)$, becomes positive and diverging towards low frequency. **(D)** The pump induced changes $\Delta\sigma_2(\omega) = \sigma_2(\omega) - \sigma_2^{eq}(\omega)$ diverge as $1/\omega$ (dotted line).

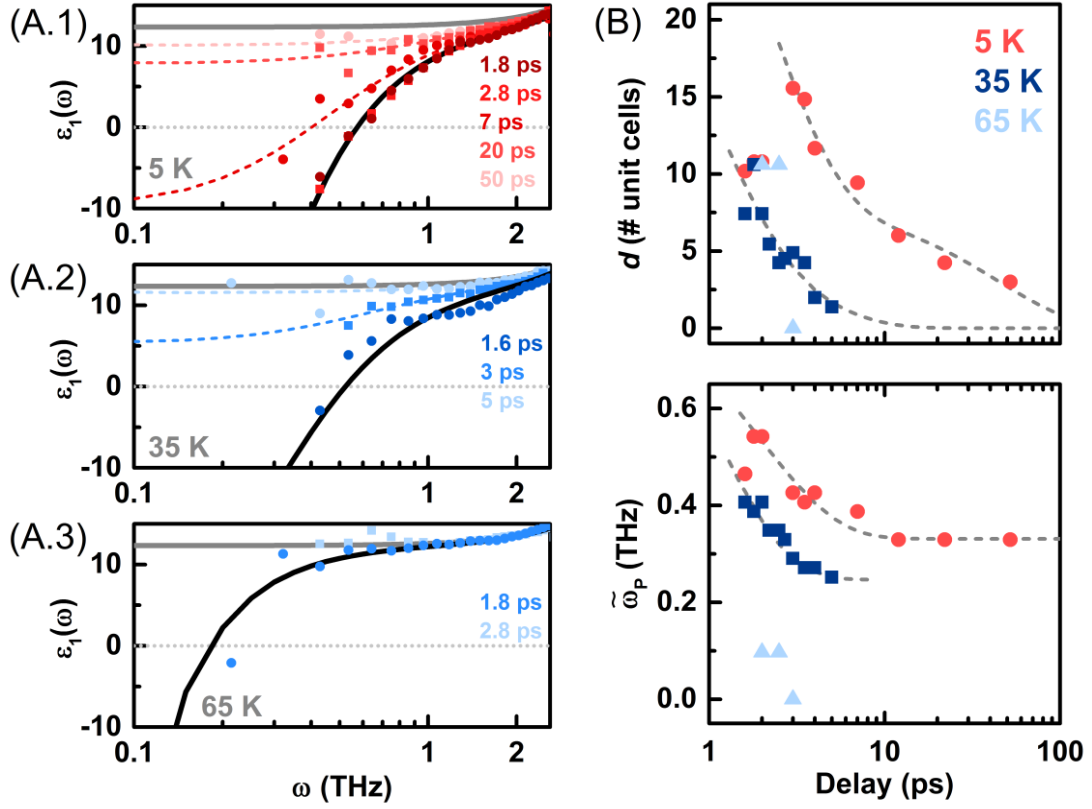

**Figure 3. Time dependence of the transient response. (A)** The real dielectric function $\varepsilon_1(\omega)$ (alternating circle and square dots) as a function of pump-probe delay at 5 K **(A.1)**, 35 K, **(A.2)**, and 65 K **(A.3)**. The response is fit with a single longitudinal mode, as described in the main text (solid black and dashed lines). The drop in coherence length causes a flattening in the low frequency $\varepsilon_1(\omega)$. **(B, top)** The c-axis coherence length of the transient plasma as a function of delay for three temperatures, 5 K (circles), 35 K (squares), and 65 K (triangles). The length is expressed in units of the $CuO_2$ plane spacing. The decay can be fit with a double exponential (dashed lines) with time constants 2 ps (1 ps) and 45 ps (4 ps) at 5 K (35 K). **(B, bottom)** The plasma frequency $\omega_P$ red shifts at early times, decaying to a constant value following single exponential (dashed lines) of 2 ps (1 ps) at 5 K (35 K).

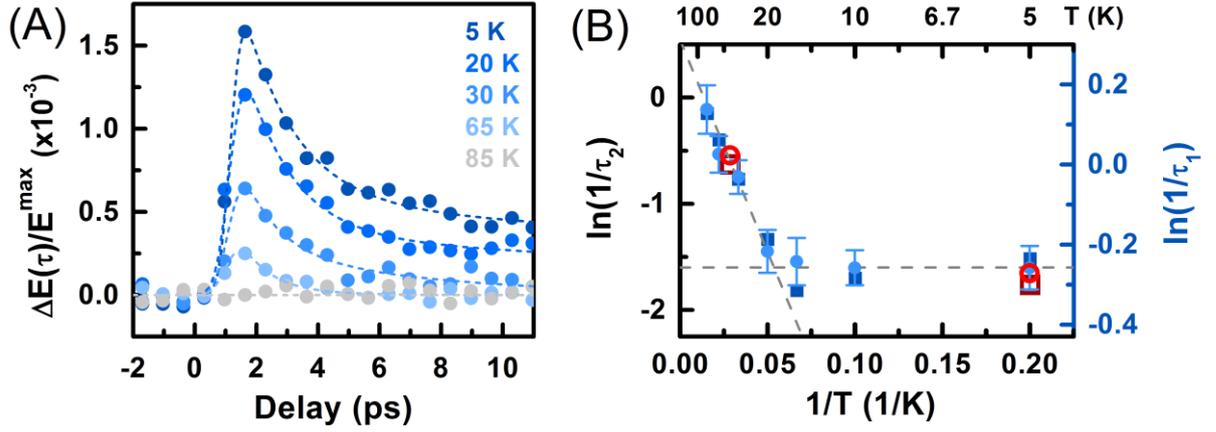

**Figure 4. Temperature dependence of the transient lifetime. (A)** The time profile of the transient state as a function of pump-probe delay. The vertical axis indicates the transient change of the peak of the THz probe field. For clarity, each point shown represents an average of 10 delay measurements. The lifetime of the transient state was extracted using a double exponential fit (dashed lines), with each exponential given roughly equal weight, 50% ± 10%. **(B)** An Arrhenius plot of the relaxation rate as a function of temperature. The short lifetime $\tau_1$ (dark blue squares) and the long lifetime $\tau_2$ (light blue circles) track the decay of the coherence length of the transient plasma. (Lifetimes used in the fits shown in figure 3C are plotted in red.) The lifetime of the transient state remains temperature independent (horizontal grey dashed line) below the spin order transition temperature, $T_{SO} \approx 25$ K. Above this transition, the lifetime exhibits an exponential temperature dependence, with an energy scale of 4 meV for $\tau_2$ and 0.8 meV for $\tau_1$ (grey dashed line).

# Supplementary: Two distinct kinetic regimes for the relaxation of light-induced superconductivity in La$_{1.675}$Eu$_{0.2}$Sr$_{0.125}$CuO$_{6.5}$

C. R. Hunt[1,2], D. Nicoletti[1], S. Kaiser[1], T. Takayama[3,4], H. Takagi[3,4,5], and A. Cavalleri[1,6]

[1] *Max Planck Institute for the Structure and Dynamics of Matter, Hamburg, Germany*

[2] *Department of Physics, University of Illinois at Urbana-Champaign, Urbana, Illinois, USA*

[3] *Department of Advanced Materials Science, University of Tokyo, Tokyo, Japan.*

[4] *Max-Planck-Institut für Festkörperforschung, Heisenbergstrasse 1, D-70569 Stuttgart, Germany*

[5] *RIKEN Advanced Science Institute, Hirosawa 2-1, Wako 351-0198, Japan*

[6] *Department of Physics, Oxford University, Clarendon Laboratory, Oxford, United Kingdom*

**Calculation of the full optical response**

Because the THz probe penetrates far deeper than the 15 μm pump, the reflected THz field includes contributions from both the photo-excited volume and the equilibrium bulk material. To isolate the optical response of the photo-excited volume alone, we model the system as a single excited layer on an unperturbed bulk (see figure S1). We set the excited layer thickness to the penetration depth of the pump intensity, $d = 85$ nm.

This model provides virtually the same results as a more elaborate multilayer model that treats the refractive index at time delay $\tau$, $\tilde{n}(\omega, \tau)$, as maximum at the sample surface, decaying exponentially with distance $z$ from the surface toward its unperturbed bulk value, $\tilde{n}_0(\omega)$, following Beer's law,

$$\tilde{n}(\omega, z) = \tilde{n}_0(\omega) + \Delta\tilde{n}(\omega)e^{-\alpha z},$$

where the linear extinction coefficient of the pump follows the inverse of the pump penetration depth, $\alpha = 1/d$.

Both models are based on defining a material's "characteristic matrix" as described in Born and Wolf.[1] Each layer $j$ in the model has a characteristic matrix of the form

$$M_j = \begin{pmatrix} \cos(k_j \tilde{n}_j \delta z) & -\dfrac{i}{p_j} \sin(k_j \tilde{n}_j \delta z) \\ -i p_j \sin(k_j \tilde{n}_j \delta z) & \cos(k_j \tilde{n}_j \delta z) \end{pmatrix}.$$

where $\tilde{n}(z) = constant$ for each layer and we define $p_j = \tilde{n}_j \cos \theta_j$ for a TE mode. The quantity $k_j$ is defined as $k_j = k_0 \cos \theta_j$, where the probe wave number in vacuum is $k_0 = \omega/c$. For the single layer model, we simply set $\delta z = d$. For the multilayer model, each layer is set to a constant $\delta z \ll d$ and the layers range from the sample surface to the probe penetration depth, $L = N \delta z$. The refractive index of the layer is defined as $\tilde{n}_j = \tilde{n}(\omega, z = j \delta z)$. The total characteristic matrix, $M$, is obtained as the product of the matrices for all layers, $M = \prod_{j=0}^{N} M_j$.

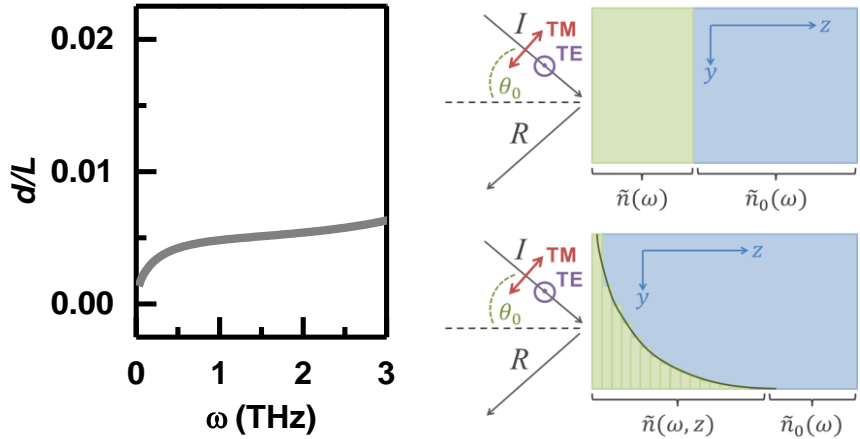

**Figure S1 Modelling the pump-probe penetration depth mismatch.** The THz probe penetrates into the sample much farther than the 15 μm pump. The ratio of the pump penetration depth $d$ to probe penetration depth $L$ in the THz range is plotted at left. To extract the surface optical response alone, we modeled the sample in three ways. Two models treat the sample as two layers—an excited film on an unperturbed bulk (top right). A third model takes into account the extinction behavior of the pump field, modeling the refractive index with an exponential dependence on depth (bottom right).

---

[1] M. Born and E. Wolf, *Principles of Optics*. (Cambridge University Press, Cambridge, 1999) 7th Edition, p. 54-64.

All measurements were taken in the TE configuration and with the THz probe oriented at $\theta_0 = 30°$ from normal incidence. From the elements of the total characteristic matrix, $m_{ij}$, we can extract the reflection coefficient,

$$\tilde{r}'(\omega) = \frac{(m_{11} + m_{12}p_L)p_0 - (m_{21} + m_{22}p_L)}{(m_{11} + m_{12}p_L)p_0 + (m_{21} + m_{22}p_L)}.$$

The quantity $p_L$ is evaluated at the probe penetration depth and $p_0 = \cos\theta_0$ is calculated at the sample surface. This equation is solved numerically for the surface refractive index $\tilde{n}(\omega, \tau) = \tilde{n}_0(\omega) + \Delta\tilde{n}(\omega, \tau)$ using a Levenberg-Marquardt fitting algorithm. The reflection coefficient is related to the measured electric field following

$$\frac{\Delta\tilde{E}(\omega, \tau)}{\tilde{E}(\omega)} = \frac{\tilde{r}'(\omega, \tau) - \tilde{r}(\omega)}{\tilde{r}(\omega)},$$

where $\tilde{r}(\omega)$ is the equilibrium complex reflection coefficient.

From the surface refractive index, we calculate the complex conductivity for a volume that is homogeneously transformed,

$$\tilde{\sigma}(\omega, \tau) = \frac{\omega}{4\pi i}[\tilde{n}(\omega, \tau)^2 - \varepsilon_\infty],$$

where $\varepsilon_\infty = 4.5$, a standard value for cuprates[2]. The optical response of LESCO 1/8 at 5 K and $\tau = 2.8$ ps is plotted in figure S2 using the single layer model (blue) and the multilayer model (red). The same data is shown in the main text in figure 2. Both models are in excellent quantitative agreement.

Because the pump penetration depth is very small compared with the probe, we considered a third model, based on an analytic approximation valid in the limit $d \ll L$. We define

$$\Delta\tilde{\sigma}(\omega, \tau) = \left(\frac{1}{Z_0 d}\right)\frac{\tilde{n}_0^2\left(1 - \frac{1}{2\tilde{n}_0^2}\right) - \frac{1}{2}}{\frac{1}{\sqrt{2}} - \tilde{n}_0\sqrt{1 - \frac{1}{2\tilde{n}_0^2}} + \sqrt{2}\left(\frac{\Delta\tilde{E}(\omega, \tau)}{\tilde{E}(\omega)}\right)^{-1}},$$

where $\Delta\tilde{\sigma}(\omega, \tau)$ is the pump induced change to the conductivity at time delay $\tau$ and $Z_0$ is the vacuum permittivity. This thin film model is also plotted in figure S2 in purple. All three models are in good quantitative agreement, with the approximate model giving a slightly enhanced

---

response. All optical properties shown in the main text were calculated with the single layer model (blue).

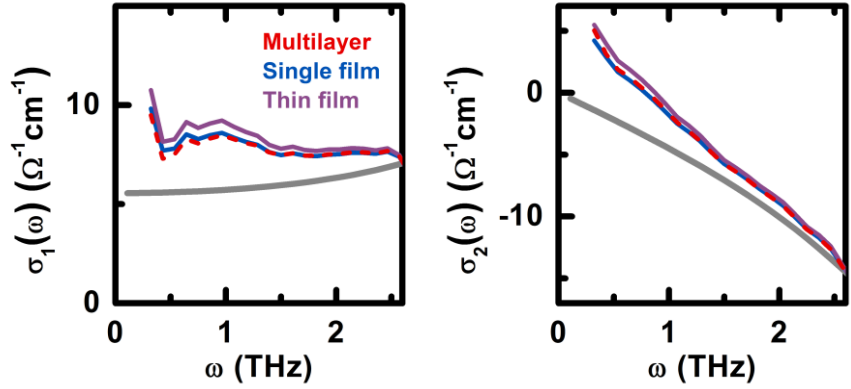

**Figure S2 The transient optical conductivity of LESCO 1/8 at 5 K calculated using three models.** The optically excited material can be well-modeled assuming a single excited layer on an unperturbed bulk (blue). A more extensive model was developed that treats the extinction of the pump into the material as an exponential decay of the transient refractive index (red). A third model, an analytic approximation assuming a large pump-probe penetration depth mismatch, is also shown for comparison (purple).

**Penetration depth mismatch**

The overall magnitude of the transient changes is sensitive to the penetration depth mismatch used by the models. However the variation in the recalculated properties when the pump penetration depth is changed by $\pm 10\%$, shown in figure S3, are well within the variation given by the three models above. Importantly, the qualitative behavior of the transient *changes* in the optical response, $\Delta\tilde{\sigma}(\omega)$, are the same regardless of the penetration depth mismatch. Figure S4 shows the same LESCO 1/8 data (blue) alongside a recalculated optical response assuming a pump penetration depth 10x larger (light blue). The data has been rescaled by a factor of 10, such that $\tilde{\sigma}(\omega) = \tilde{\sigma}_0(\omega) + 10\,\Delta\tilde{\sigma}(\omega)$, to illustrate that the transient changes retain the same qualitative behavior, namely, the diverging $\Delta\sigma_2$ which is characteristic of a high mobility state and a London-like superconducting response.

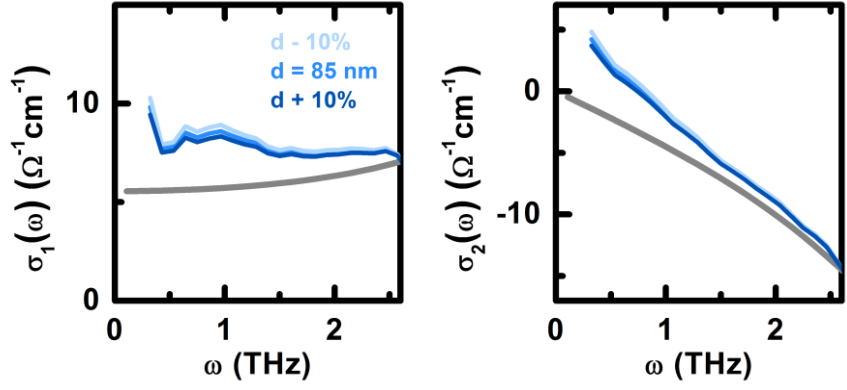

**Figure S3 Effect of varying the penetration depth.** The penetration depth is determined experimentally by the in-plane optical response at 15 μm. However varying the penetration depth by some fraction does not significantly change the transient response. Here, the penetration depth $d$ is varied by $\pm 10\%$, such that $d' = d \pm 0.1d$.

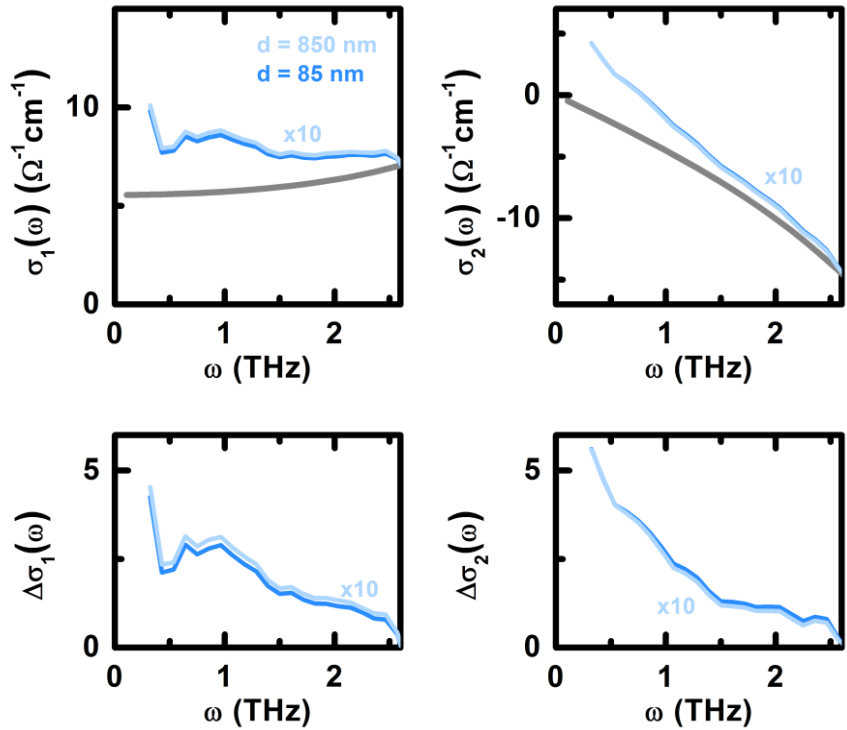

**Figure S4 Scaling of the transient response with penetration depth.** The magnitude of the pump-induced changes in the transient response scale with the penetration depth mismatch. However, the qualitative behavior of the changes which mark the generation of a high mobility state, including a diverging $\Delta\sigma_2$, are present regardless of the magnitude of the mismatch.

**Full Optical Response in the Charge Ordered Regime**

The transient optical response is qualitatively the same throughout the charge order regime. The full optical properties at several temperatures up to 65 K are shown in Figure S5. The plasma mode crosses zero at the limit of our spectral resolution at 65 K. Based on the temperature dependence of the optical response, we project that the mode survives to the charge order transition, 80 K.

The fits (dashed lines) are of the form that is described in the main text, with the addition of an additional Lorentzian term to account for contributions to $\sigma_1(\omega)$ not related to the plasma mode. The addition of this term does not affect the best fit values of the plasma frequency $\omega_p$ or scattering rate $\Gamma$. The values of $\omega_p$ and $\Gamma$ are uniquely determined by taking into account the full complex response. As a technical comment, although the most dramatic consequence of $\Gamma$—the frequency of the characteristic peak in $\sigma_2(\omega)$ and low frequency increase in $\sigma_1(\omega)$—are at the edge of our frequency resolution at earliest time delays, the value of $\Gamma$ plays a role in setting the shape of the response throughout our entire frequency window.

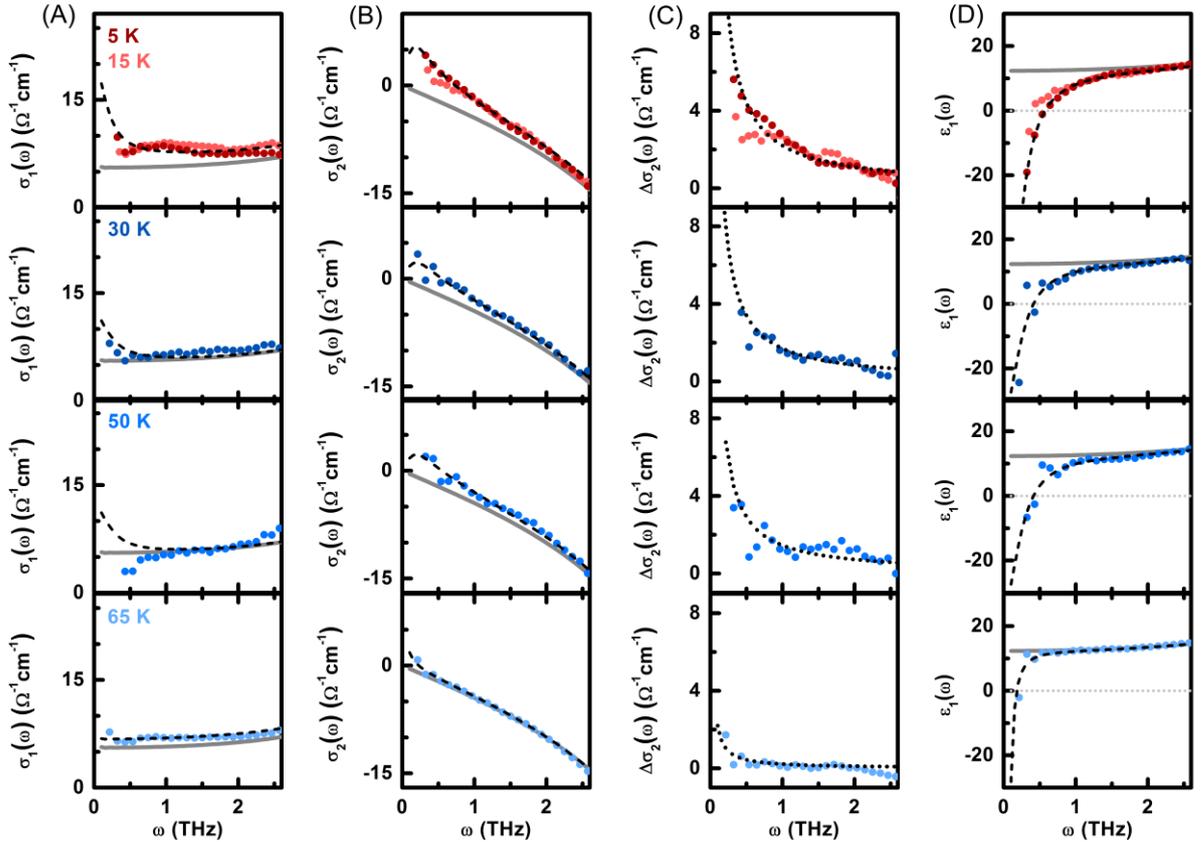

**Figure S5 Full optical response $\sigma_1(\omega)$ (A) and $\sigma_2(\omega)$ (B) at 5 temperatures.** The optical response is characterized by a zero crossing in $\varepsilon_1(\omega)$ **(D)** and corresponding diverging behavior in pump-induced changes in the inductive conductivity $\Delta\sigma_2(\omega)$ **(C)**.